\documentstyle{mn}
\input{epsf}

\def\ergs{~{\rm erg}~{\rm s}^{-1} }
\def\ergscm2{~{\rm erg}~{\rm s}^{-1}~{\rm cm}^{-2} }
\def\MeV{~\rm{MeV}}
\def\keV{~\rm{keV}}
\def\eV{~\rm{eV}}
\def\G{~\rm{G}}
\def\s{~\rm{s}}
\def\cm{~\rm{cm}}
\def\Lg{$L_\gamma$}
\def\Lx{$L_X$}
\def\edot{$L_{\rm sd}$}
\newbox\grsign \setbox\grsign=\hbox{$>$} \newdimen\grdimen \grdimen=\ht\grsign
\newbox\simlessbox \newbox\simgreatbox \newbox\simpropbox
\setbox\simgreatbox=\hbox{\raise.5ex\hbox{$>$}\llap
     {\lower.5ex\hbox{$\sim$}}}\ht1=\grdimen\dp1=0pt
\setbox\simlessbox=\hbox{\raise.5ex\hbox{$<$}\llap
     {\lower.5ex\hbox{$\sim$}}}\ht2=\grdimen\dp2=0pt
\setbox\simpropbox=\hbox{\raise.5ex\hbox{$\propto$}\llap
     {\lower.5ex\hbox{$\sim$}}}\ht2=\grdimen\dp2=0pt
\def\simgreat{\mathrel{\copy\simgreatbox}}
\def\simless{\mathrel{\copy\simlessbox}}

\title[High-energy emission]
{High-energy emission from pulsars in polar-cap models
with CR-induced cascades}
\author[B. Rudak \& J. Dyks]
	{B. Rudak \& J. Dyks \\
N. Copernicus Astronomical Center, Rabia\'nska 8, 87-100 Toru\'n,
Poland\\
e-mail: bronek@camk.edu.pl, jinx@astri.uni.torun.pl  }

\date{Accepted 1998 November 30. Received 1998 June 2; in original form 1997 March 4}

\begin{document}

\maketitle

\label{firstpage}

\begin{abstract}

For a subclass of polar-cap models based on electromagnetic cascades induced
by curvature radiation (CR)
we calculate
broad-band high-energy spectra of pulsed emission expected for classical
and millisecond pulsars. 
The spectra are a combination of curvature and 
synchrotron components.
The spectrum of
curvature component breaks at $150 \MeV$,
and neither its slope nor level below this energy are compatible with phase-averaged
spectra of pulsed X-ray emission inferred from observations.
Spectral properties in the combined energy range 
of {\it ROSAT} and {\it ASCA}
(0.1 - 10 keV) depend
upon the location of cyclotron turnover energy 
$\epsilon_{\rm ct} = \hbar {e B \over m_{\rm e} c} /\sin \psi$ in the synchrotron component. 
Unlike in outer-gap models, the available
range of pitch angles $\psi$ is rather narrow and confined to low values.
For classical pulsars, a gradual turnover begins already 
at $\sim 1 \MeV$, and
the level of the synchrotron spectrum
decreases. At $\sim 10 \keV$ the 
curvature component eventually takes over, but with
photon index
$\alpha = 2/3$, 
in disagreement with observations. \hfill\break
For millisecond pulsars, the X-ray spectra are dominated
by synchrotron component with
$\alpha \simeq 1.5$, and a sharp turnover into
$\alpha \simeq -1$ at $\epsilon_{\rm ct} \sim 100 \eV$. \hfill\break
Relations of pulsed luminosity $L_{\rm X}$ to
spin-down luminosity \edot~ are presented for classical and millisecond pulsars.
We conclude that 
spectral properties and fluxes of pulsed non-thermal X-ray
emission of some objects, like the Crab or
the millisecond pulsar B1821-24, pose a challenge to 
the subclass of polar-cap models based on curvature and 
synchrotron radiation alone.

\end{abstract}


\section{Introduction}
According to recent results from {\it ROSAT} and {\it ASCA} (Becker \& Tr{\"u}mper 1997 and
Saito 1998, respectively) 
luminosities of pulsed and unpulsed components in the X-ray emission from pulsars
are related to spin-down luminosities $L_{\rm sd}$, suggesting
thus a rotation-powered origin of X-rays. 
The unpulsed emission is usually interpreted as synchrotron radiation from
an unresolved nebula surrounding the pulsar. A magnetospheric wind of ultrarelativistic
particles (with Lorentz factors about $10^7$) will lead to X-ray emission
provided the nebula contains magnetic field  of the order of $\sim 10^{-4}\G$ (e.g.
Manning \& Willmore (1994) estimate the magnetic field strength of $0.17 \times 10^{-4}\G$ for the hypothetical
nebula around B0950+08). 

Pulsed components are of particular interest, since
they are the direct signatures of non-thermal processes within a magnetosphere.
Very likely, pulsed non-thermal X-ray
emission is nothing but a low-energy tail of
gamma-ray emission, which in most models 
is a superposition of
curvature (CR) and synchrotron (SR) emission. In a subclass of polar
cap models developed by Dermer \& Sturner (1994) gamma-ray photons
have different origin - they are thermal UV/X-ray photons subject to Lorentz-boost 
by inverse Compton scattering (IC) with beam particles. 

In polar-cap scenarios pulses in gamma-rays should
be accompanied by pulses in X-rays, with
similar shapes, and in phase. With such criteria the Crab pulsar would be a model example.
However, one has to show first that its contributions to X-rays and gamma-rays
are in right proportions, i.e. in (at least qualitative) accord with model predictions.
 
Phase-averaged broad-band ($X\gamma$) spectra of six, out of seven, gamma-ray pulsars
are relatively steep, with photon index $\alpha_{X\gamma}$ decreasing with
pulsar's characteristic age (see Thompson et al. 1997 for recent review):\hfill\break 
For the Crab $\alpha_{X\gamma} \approx 2$,
and therefore,
luminosities per logarithmic energy bandwidth in X-rays and in gamma-rays are comparable.
For B1055-52, which is the oldest gamma-ray pulsar, $\alpha_{X\gamma} \approx 1.6$, and
its pulsed X-ray luminosity \Lx~ is roughly one per cent of its \Lg. \hfill\break
[Note: To make this comparison we followed the convention used in gamma-rays 
($\Omega_\gamma = 1$~sr is the solid angle of gamma-ray emission)
taking $\Omega_X = 1$~sr for the solid angle of X-ray emission.]

These observed spectral properties have been quite successfully accounted for
in a newly proposed model of `thick outer gap' by Zhang \& Cheng (1997).
Unlike polar-cap models, outer-gap models [see also Wang et al. (1997)]
may
take advantage of the full range of pitch angles, which
in turn makes it possible to trigger electromagnetic cascades even
for relatively soft gamma-ray photons 
($\simgreat 1 \MeV$) on magnetic field lines (also on closed ones). The energy distribution
function for e$^\pm$ pairs produced in the cascades may easily become steep enough
($N(\gamma_\pm) \propto \gamma_\pm^{-3}$) over a wide range of energy,
for subsequent synchrotron spectrum ($\alpha_{X\gamma} = 2$)
to account for the observed soft gamma-rays and X-rays. In particular, Cheng et al. (1998)
managed to reconstruct simple empirical relation between \Lx~ and \edot, which reads
(assuming $\Omega_X = 4\pi$~sr) $L_X \simeq 10^{-3}L_{\rm sd}$
(Becker \& Tr{\"u}mper~1997).
Moreover, cascades propagating starward
are subject to substantial lateral spread, leading thus to rather weak X-ray modulations. 

The aim of this paper is to present general features of broad-band $X\gamma$ spectra
expected for the subclass of polar cap models, with $X\gamma$ emission due to CR and SR
(e.g. Daugherty \& Harding 1982, Daugherty \& Harding 1996) and to show that
luminosity of pulsed X-rays observed for some pulsars is too high 
compared to \Lg~ (or to \edot~ if there
is no information about gamma-rays) to be understood within
polar-cap models unless some unorthodox assumptions are accepted.
Some of these features are only weakly model-dependent and may, therefore, play
a decisive role in asessing validity of polar-cap models.
We were motivated by the fact, that neither proponents nor oponents of the polar-cap models
seem to acknowledge a real challenge posed to these models by observed properties of
non-thermal, pulsed X-ray emission. It is this emission which  
is more promising as a potential discriminator between rivalring classes (polar-cap, outer-gap) of
models than gamma-ray emission itself.\hfill\break 
In Section 2 we present 
model spectra of CR and SR emission per parent (beam) particle, expected in pulsars with
high ($\sim 10^{12}\G$) and low ($\sim 10^{9}\G$) magnetic fields. 
Section 3
compares \Lx~ expected within energy band $0.1\keV - 10 \keV$ with
observations
for
a wide range of spin-down luminosities \edot.
Section 4 contains discussion and conclusions.

\section{Spectral Properties of Curvature and Synchrotron Radiation}

The energy distribution $N_{\rm e}(\gamma)$ of particles is governed by their injection 
rate $Q_{\rm e}(\gamma)$ (also called a source function), 
cooling rate $\dot \gamma$, and a characteristic time scale of their escape $t_{\rm esc}$, via 
steady-state kinetic equation
\begin{equation}
{\partial \over {\partial \gamma}}(|\dot \gamma| N_{\rm e}) = 
                                               N_{\rm e}/ t_{\rm esc} - Q_{\rm e},
\label{kin0}                                               
\end{equation}
where $\gamma$ is particle energy in units of $m_{\rm e}c^2$.

The solution of eq.\ref{kin0} yields two asymptotic relations
\begin{equation}
N_{\rm e}(\gamma) = \cases{
          {1 \over {|\dot \gamma|}} \int_\gamma^{\gamma_{\rm max}}
           Q_{\rm e}(\gamma) d \gamma,&$|\dot \gamma| / \gamma \gg 
           t_{\rm esc}^{-1}$; \cr
          Q_{\rm e} \cdot t_{\rm esc},&$|\dot \gamma| / \gamma \ll 
           t_{\rm esc}^{-1}$. \cr}
\label{kin1}
\end{equation}
If the solution to eq.\ref{kin0} is a power-law function
with index $p$, $N_{\rm e} \propto \gamma^{-p}$, 
the subsequent photon spectrum of SR or CR will have a form
\begin{equation}
N_\nu(\epsilon) \propto \epsilon^{-{p + 1 \over q}},
\label{e0}
\end{equation}
(but the value of $p$ must not be too low) where $q = 2$ for SR, 
or $q = 3$ for CR. \hfill\break
[Note: Throughout this section we use a short notation
for electron and photon distribution functions: $N_{\rm e}(\gamma) = dN_{\rm e}/d\gamma$ and
$N_\nu(\epsilon) = dN_\nu/d\epsilon$, respectively.]

\subsection{Curvature Radiation}
First, let us consider spectral component due to CR.
If the source function $Q_{\rm e}(\gamma)$ of beam particles is monoenergetic
\begin{equation}
Q_{\rm e}(\gamma) = C_0 \, \delta (\gamma_0 - \gamma),
\label{Q}
\end{equation}
and redistribution over the energy space is governed by the CR cooling
($|\dot \gamma| = |\dot \gamma_{\rm cr}| \propto \gamma^4$), then from the first relation
of eq.\ref{kin1} it follows that $N_{\rm e} \propto \gamma^{-4}$ (i.e. $p = 4$) for
$\gamma < \gamma_0$. The lower limit (let us denote it  as $\gamma_{\rm break}$) to this distribution
is determined by the condition, that a cooling time-scale due to CR,
$t_{\rm cr} \equiv \gamma/ |\dot \gamma_{\rm cr}|$, is shorter than $t_{\rm esc}$.
According to eq.\ref{e0}, 
the photon spectrum of CR becomes $N_{\rm cr}(\epsilon) 
\propto \epsilon^{-{5 \over 3}}$. It remains unchanged down to photon energy $\epsilon_{\rm break}$,
which may be approximated with a characteristic energy of CR
\begin{equation}
\epsilon_{\rm crit} = {3 \over 2} \, c \, \hbar \, {\gamma^3 \over \rho_{\rm cr}},
\label{e1}
\end{equation}
(where $\rho_{\rm cr}$ is a local radius of curvature)
taken for $\gamma_{\rm break}$.
Below $\epsilon_{\rm break}$, 
the photon spectrum follows 
the low-energy tail of CR, i.e. $N_{\rm cr}(\epsilon) \propto \epsilon^{-{2 \over 3}}$.
In order to find $\epsilon_{\rm break}$ we should first estimate the time scale $t_{\rm esc}$. 
We assume that  $t_{\rm esc} \approx \rho_{\rm cr}/ c$.
The CR cooling rate for a single particle of energy $\gamma$ is
\begin{equation}
\dot \gamma_{\rm cr} = - {2 \over 3} \, {e^2 \over m c}\, {\gamma^4 \over \rho_{\rm cr}^2}.
\end{equation}
The condition $t_{\rm cr} = t_{\rm esc}$ becomes then 
\begin{equation}
{\gamma_{\rm break}^3 \over \rho_{\rm cr}} = {3 \over 2} \, {m c^2 \over e^2},
\label{e2}
\end{equation}
and from eq.\ref{e1} it follows that 
\begin{equation}
\epsilon_{\rm break} = {9 \over 4} \, \hbar \, {c \over r_0} \approx 150 \MeV,
\label{e3}
\end{equation}
where $r_0$ is the classical electron radius. Note, that 
the photon energy $\epsilon_{\rm break}$ at which the spectral break occurs does not depend on any
pulsar parameters, as long as our estimation of  $t_{\rm esc}$ is accurate. In particular, it
does not depend on magnetic field structure - should it be pure dipole with $\rho_{\rm cr}
\simgreat 10^8 \sqrt{P} \cm$, or dominated by high-order multipolar components with
$\rho_{\rm cr} \sim 10^6\cm$.

Fig.1 shows numerically calculated shapes of the CR spectrum produced by beam particles injected 
at the outer rim of a canonical polar cap with a dipolar magnetic field. The magnetic
strength $B_{\rm pc}$ at the polar cap is $10^9\G$ (upper panel), and $10^{12}\G$ (lower panel).
The period of rotation $P$ is $0.003 \s$ and $0.06 \s$, respectively. For the
initial energy $E_0$ we choose $1.1 \times 10^7\MeV$ and $6.7\times 10^6\MeV$, respectively (see section 3
for explanation). 
The dotted line corresponds to the unabsorbed spectrum, and the
solid line is the spectrum with magnetic absorption.
The spectral break (where the power-law spectrum changes its slope by one power of $\epsilon$) 
is visible in both panels, and its location 
at $\sim 150 \MeV$ agrees quite well with the analytical estimate (eq.\ref{e3}). A word of comment on  
high-energy 
cutoffs seems appropriate here, though it is not linked directly to the subject of this paper:
The high-energy cutoff for the case of $10^{12}\G$ does not exceed $10$~GeV, whereas
the cutoff for the low-$B$ case reaches $\sim 0.1$ TeV (the energy range accessible
with ground-based Cherenkov techniques). The explanation is twofold:
(1) 
for low values of $B$ the magnetic absorption is weak; (2) very short periods $P$ infer 
small  
dipolar curvature radii, and a characteristic photon energy (eq.\ref{e1}) taken for
$\gamma_0 = E_0/m_{\rm e}c^2$ increases. 

\begin{figure*}
\begin{center}
\leavevmode
\epsfxsize=16 cm 
\epsfbox{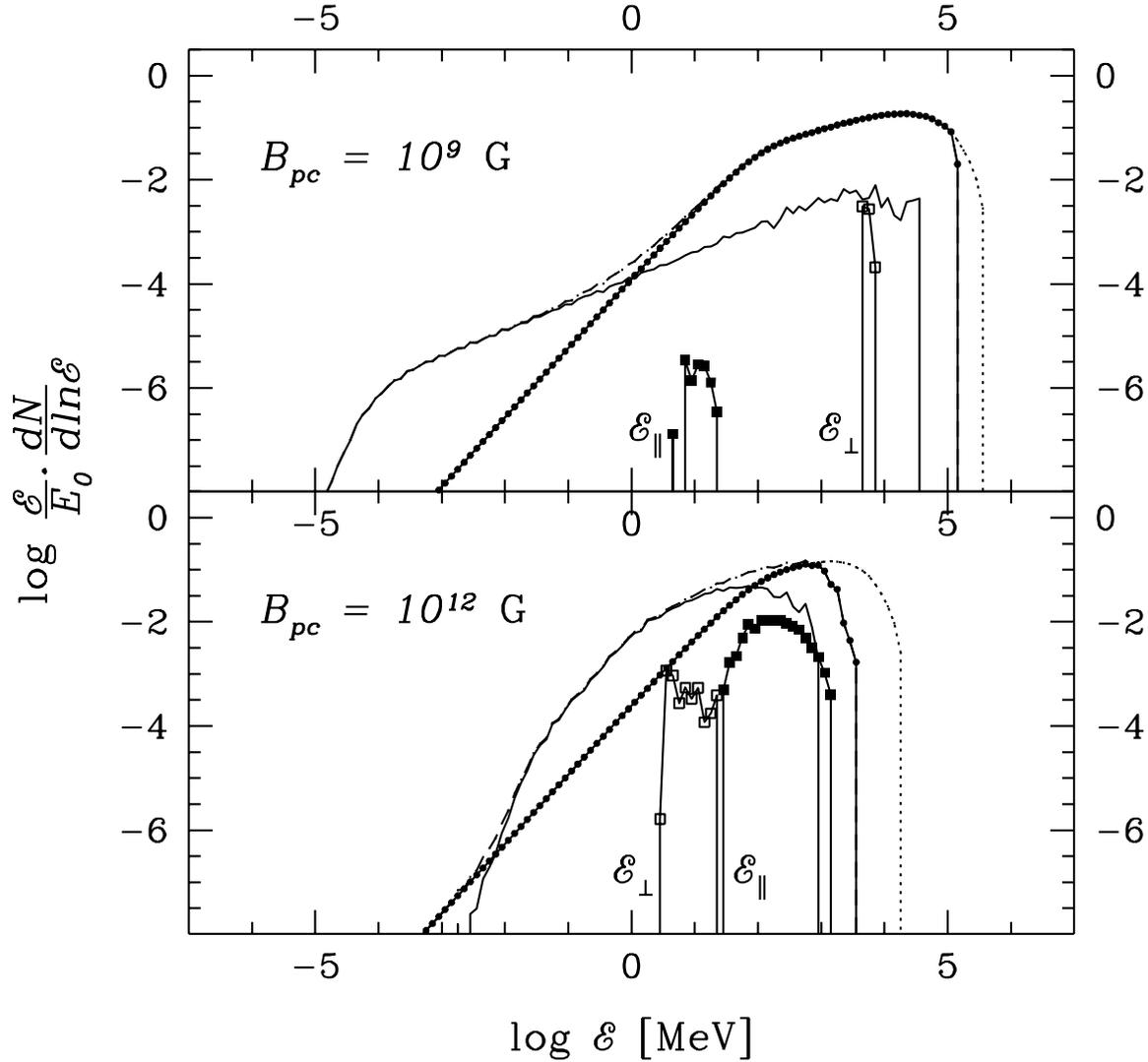}
\end{center}
\caption{
The radiation energy spectrum per logarithmic energy bandwidth.
The spectrum is 
normalized to the energy of the parent particle, $E_0$. It consists of 
two components: curvature and synchrotron.
Dotted line is the curvature spectrum before correction for the magnetic absorption
effects. Solid line connecting filled dots, is the curvature spectrum with the magnetic absorption
taken into account. Solid line without any symbols overlayed is the synchrotron component.
Dot-dashed line indicates superposition of the two. \hfill\break
In addition to the electromagnetic spectra, we show the spectra of e$^\pm$ pairs: the line
connecting filled squares is for the energy parallel to local magnetic field lines
(labeled with ${\cal E}_\parallel$);
the line connecting open squares is the initial distribution of energy perpendicular
to local magnetic lines (labeled with ${\cal E}_\perp$).\hfill\break
The upper panel is for $B_{\rm pc}=10^{9}\G$, and 
$E_0 = 1.08 \times 10^7 \MeV$ (see eq.\ref{E0}) which corresponds to $P = 3.1\times 10^{-3}\s$
(i.e. $L_{\rm sd} = 10^{35}\ergs$).\hfill\break
The lower panel is for $B_{\rm pc}=10^{12}\G$, and
$E_0 = 6.68 \times 10^6 \MeV$ which corresponds to $P = 5.6\times 10^{-2}\s$
(i.e. $L_{\rm sd} = 10^{36}\ergs$)
}
\label{fig:fig1}
\end{figure*}

With the spectral break at $\epsilon_{\rm break} \approx 150 \MeV$, the  
curvature radiation becomes energetically unimportant below this energy. Such objects 
would be dim in X-rays, which apparently is not the case.
Moreover, the
slope $-2/3$ in the X-ray energy range is in clear conflict with spectral analysis of
X-ray data (Becker \& Tr\"umper 1997).

A way out may be offered by including synchrotron radiation (next subsection)
and/or by relaxing the assumption about the monoenergetic source function of
beam particles (eq.\ref{Q}). Suppose, for example, that we want the slope
of $-5/3$ to extend 
down to energy $\epsilon_x$ in the X-ray range:
$N_\nu(\epsilon) \propto \epsilon^{-{5 \over 3}}$ for  $\epsilon \simgreat \epsilon_x$.
According to eq.\ref{kin1} (lower relation)
we now have 
the energy distribution $N_{\rm e}(\gamma) = Q_{\rm e} \cdot t_{\rm esc}$.
This, in turn, requires that the source function of beam particles is
\begin{equation}
Q_{\rm e} \propto \gamma^{-4},\,\, {\rm for} \,\,\gamma \simgreat \gamma_x,
\label{Q1}
\end{equation}
a very steep function of $\gamma$ (for whatever the reason), where 
$\gamma_x$ is linked to
$\epsilon_x$ via eq.\ref{e1}.
To find $\gamma_x$ we
solve the equation $\epsilon_x = \epsilon_{\rm crit}(\gamma_x)$
for $\gamma_x$:  
For dipolar magnetic fields, the smallest
curvature radii $\rho_{\rm cr}$ are those at outer rim of the polar cap. They may be approximated with
$\rho_{\rm cr} \approx \sqrt{r \, R_L}$, where $r$ is the radial coordinate of the particle, and 
$R_L = c \, P /2\pi$ is
the pulsar's light cylinder. Within the region of $r \simeq 2 R_{NS}$ we have therefore
$t_{\rm esc} \approx 10^{-2.5} \sqrt{P}\s$. From eq.\ref{e1} we obtain 
\begin{equation}
\epsilon_x \approx 0.3\, (\gamma_x/10^5)^3 \, P^{-0.5} \keV.
\end{equation}
For a pulsar with $P = 0.1 \s$, the source function $Q_{\rm e}(\gamma) \propto \gamma^{-4}$ would have
to extend down to $\gamma_x \sim 10^5$ for $\epsilon_x$ to reach $\sim 1 \keV$. This value of
$\gamma_x$ is
about 55 times lower than the value of $\gamma_{\rm break}$ (eq.\ref{e2})
corresponding to $\epsilon_{\rm break}$ (eq.\ref{e3}).

\subsection{Synchrotron Radiation}
Since CR is not a promising candidate for hard X-rays, the standard explanation 
involves SR, produced by e$^\pm$ pairs created in the process of magnetic absorption
of high-energy CR photons. Total contribution of the SR component to 
the overall (SR+CR) energy of radiation per beam particle depends on number of created pairs $n_\pm$.
One may expect, therefore, that at least for pulsars with strong magnetic fields ($\sim 10^{12}\G$),
the energy contained in SR should be comparable or exceed the energy contained in CR. 
However, spectral properties of the SR component depend not only on local values of $B$,
but also on the available
range of pitch angles $\psi$ between magnetic field lines and the direction of
propagation of created pairs. The most popular assumption -- about an isotropic
distribution -- may be fully justified in the case of outer-gap models (Zhang \& Cheng 1997),
but is not obvious by any means for polar-cap scenarios. 

Below, we'll present numerical
calculations of synchrotron spectra due to e$^\pm$-pairs which are assumed 
(after Daugherty \& Harding 1982) to be directed
along the direction of propagation of their parent
CR-photons at the moment of the creation. 
This assumption, along with narrow opening
angles of magnetic field lines available for beam particles, results in 
a confinement of pitch angles to a narrow range of low values.  
For a pulsar with a dipolar magnetic field and rotating with period $P$, we
expect then
$\sin \psi \leq 0.01/\sqrt P$
as a first approximation.

The character of the source function $Q_\pm$ of e$^\pm$-pairs depends primarily on the
reachness of the cascades. In the case of low magnetic fields ($\sim 10^{9}\G$)
only one generation of e$^\pm$-pairs is produced, and their formation region is narrow.
For these reasons $Q_\pm$ is almost monoenergetic (see Fig.1, upper panel), and basic spectral
properties of SR calculated for a canonical millisecond pulsar are easily understood
with well known analytical considerations:

Let us denote by $\gamma_\parallel$ the particle Lorentz factor along the local magnetic 
field line. We will ignore possible curvature of the line, and assume that $\gamma_\parallel$
remains constant. The particle energy (in units of $m_{\rm e}c^2$) to be emitted in bursts of SR corresponds
then to
Lorentz factor $\gamma_\perp$ of gyration. The total particle energy
is $\gamma = \gamma_\perp \, \gamma_\parallel$. As long as $\gamma_\parallel \gg 1$, 
the pitch angle of the particle is determined by $\sin \psi \approx \gamma_\parallel^{-1}$.

The rate of SR cooling is
\begin{equation}
\dot \gamma_{\rm sr} = - {2 \over 3} \, 
{r_0^2 \over m_{\rm e} c}\, B^2 \gamma_\perp^2 = - {2 \over 3} \,
{r_0^2 \over m_{\rm e} c}\, B^2 \sin ^2\psi \,\gamma^2.
\label{SR0}
\end{equation}
Unlike in the case of CR, the SR cooling rate is enormous and affects the energy 
distribution of pairs 
until $\gamma_\perp \sim 1$ where the synchrotron approximation breaks (O'Dell \& Sartori 1970).
Therefore, from eqs.\ref{kin1}, \ref{SR0} and \ref{e0}, we'll get photon spectrum of SR
with well known single power-law shape
$N_{\rm sr}(\epsilon) \propto \epsilon^{-{3 \over 2}}$. 
Main contribution to the spectrum at energy $\epsilon$ comes from particles with 
$\gamma$ for which $\epsilon = \epsilon_{\rm sr}(\gamma)$, where
\begin{equation}
\epsilon_{\rm sr} = {3 \over 2} \, c \, \hbar \, {e B \over m_{\rm e} c} \gamma^2 \sin \psi,
\label{SR1}
\end{equation}
is a critical photon energy in SR.  
The spectrum spreads between a high-energy
cutoff
$\epsilon_{\rm sr}(\gamma_{\rm max})$ and 
a low-energy turnover $\epsilon_{\rm ct}$ determined by the condition $\gamma_\perp \sim 1$
(O'Dell \& Sartori 1970):
\begin{equation}
\epsilon_{\rm ct} \equiv \epsilon_{\rm sr}(\gamma = \gamma_\parallel) 
= {3 \over 2} \, c \, \hbar \, {e B \over m_{\rm e} c} \,{1 \over {\sin \psi}}.
\label{SR2}
\end{equation}
Below $\epsilon_{\rm ct}$, the spectrum 
flattens, and may be described asymptotically as
$N_{\rm sr}(\epsilon) \propto \epsilon^{+1}$.
It is built up by contributions from low-energy tails emitted by particles with $\gamma_\perp \gg 1$, 
and each low-energy tail is assumed to cut off at local gyrofrequency, which in the reference frame
comoving with the center of gyration is $\omega_B = {e B \over m_{\rm e} c \,\gamma_\perp}$.
 
The numerical example presented in the upper panel of Fig.1 
reveals all these `classical textbook' features so clearly, because (for reasons mentioned
before) the source function of
pairs in the case of a dipolar field
with $B_{\rm pc}= 10^9\G$ is practically monoenergetic. 
The turnover energy 
is located around $0.1 \keV$, i.e. in the soft range of X-rays. More importantly, the SR component
dominates over the CR component in the entire energy range of X-rays.
The lower panel of Fig.1 shows analogous results obtained for $B_{\rm pc}= 10^{12}\G$.
The created pairs belong now to two generations, and
they are rich ($\sim 400$ times more numerous than for $10^9\G$). Therefore, the SR component
is now energetically comparable to CR. 
There is no single, well defined turnover energy  
$\epsilon_{\rm ct}$ anymore. The spectrum of the source function $Q_\pm$ is spread
over two decades in energy, and in consequence
the SR spectrum reveals a gradual turnover, which starts already at $\sim 1\MeV$, due to
high values of $\gamma_\parallel$ as well as strong local $B$ (see eq.\ref{SR2}).

Our approximate treatment of SR spectral shapes at low-energy limit requires a word of comment:
As electrons go to the ground Landau level,
the spectrum reveals its harmonic structure and the analytical formula that we used,
should be treated with caution.
Comparison of Monte Carlo spectra (calculated by summing-up the rates of the quantum transitions), 
with
analytical formulae was carried out by Harding \& Preece (1987). They found that
the analytical formula for the low-energy tail of SR with a cutoff at cyclotron frequency
generally
overestimates (but not dramatically)
the actual level of photon spectrum. Therefore, our results for SR in the context
of X-rays should be treated rather as upper limits, but this does not change our conclusions.

\section{Relations Between X-rays and Gamma-rays}
We calculated combined spectra 
of CR and SR (as shown in Fig.1), emitted by a beam particle with initial energy $E_0$,
for two canonical
values of $B_{\rm pc}: \, 10^9\G$ and $10^{12}\G$, and for a range of periods $P$ in order
to cover a full possible range of spin-down luminosities \edot.
Formula for the energy $E_0$ was adopted from Rudak \& Dyks (1998) [RD98]. This energy is only a 
few times higher than threshold energy $E_{\rm min}$ 
required for magnetic pair creation in a dipolar magnetic field:
\begin{equation}
E_0 = 2.5 \times E_{\rm min}, 
\label{E0} 
\end{equation}
\begin{equation}
{\rm with} \>\, 
E_{\rm min} = 1.2 \times 10^7 \, \left({B \over 10^{12}\G}\right)^{-1/3} P^{1/3}\MeV,
\end{equation}
but at the same time it must never exceed any of the following limits -- $E_{\rm max}$
and $E_{\rm w}$ (the first restriction is due to curvature cooling, the second one is
due to potential drop across the polar cap). 
The original motivation for this formula
was to reproduce
\Lg~ for the seven gamma-ray pulsars. For $10^9\G$, $E_0 \approx 10^7\MeV$, while 
\edot~ covers the range between $10^{34}\ergs$ and $10^{37}\ergs$. In practice, the values
of $E_0$ do not change dramatically over the allowed ranges of \edot:
For $10^{12}\G$, $E_0$ starts with $3\times 10^6\MeV$, increases to $9.8\times 10^6\MeV$,
and goes down to $4\times 10^6\MeV$, for 
\edot~ changing from $10^{31}\ergs$ to $10^{39}\ergs$.

What became clear already in the previous section
is that the range of X-rays is never energetically important. Although for
$10^9\G$ the spectrum of SR extends well into soft X-ray band and 
dominates there,
its fractional contribution to the total radiation energy output is low.
It varies between $0.003$ (for $L_{\rm sd} = 10^{34}\ergs$) and 0.22 (for $10^{37}\ergs$).
For the case of 
$10^{12}\G$, the energy content of SR becomes significant for 
$L_{\rm sd} > 10^{33}\ergs$,
but the spectrum of SR is now confined to gamma-rays, and it turns over at $\simless 1 \MeV$.
In either case, the bulk of particle energy converted into radiation
is concentrated within gamma-rays. 

Gamma-ray luminosity \Lg~ in RD98 was identified with the power of outflowing particles 
\begin{equation}
L_{particles} = \eta \, E_{\pm} n_{\pm} \dot N_{\rm GJ},
\label{Lp}
\end{equation}
where $E_{\pm}$ (characteristic energy attained by a fraction of secondary particles,
$\eta \, n_{\pm}$, due to 
possible acceleration) was
assumed to be of the order of $E_0$, and  
$\dot N_{\rm GJ}$ was the Goldreich-Julian rate of outflow of beam particles.
The parameter $\eta = 0.004$ was used to match the model with the joint {\it EGRET} and {\it COMPTEL} 
luminosity inferred for B1951+32 (see RD98 for detalis).
Here we made more accurate calculations for \Lg, by replacing the particle energy $E_0$ with its 
radiation energy yield $E_\gamma$ for photon energy $\epsilon \geq 100 \keV$.
Similarly, we calculate expected X-ray luminosity 
\Lx~ within $(0.1\keV - 10\keV)$. 
Accordingly, we take    
\begin{equation}
L_\gamma =  L_{particles} \times {E_\gamma \over E_0}, \>\, 
{\rm and} \>\, L_X =  L_{particles} \times {E_X \over E_0},
\label{Lgx} 
\end{equation}
where ${E_\gamma \over E_0}$ and ${E_X \over E_0}$ are fractional radiation energy yields per particle,
calculated by integrating the differential (per logarithmic energy bandwidth) spectra 
[which are simply
\begin{equation}
{1 \over E_0} \epsilon\,{d N_\nu \over d \ln \epsilon}
\label{yield} 
\end{equation}
-- see Fig.1 for two examples],
over the range $\epsilon \geq 100 \keV$ and $0.1\keV \leq \epsilon \leq 10 \keV$, respectively.

\begin{figure}
\begin{center}
\leavevmode
\epsfxsize=8.5 cm 
\epsfbox{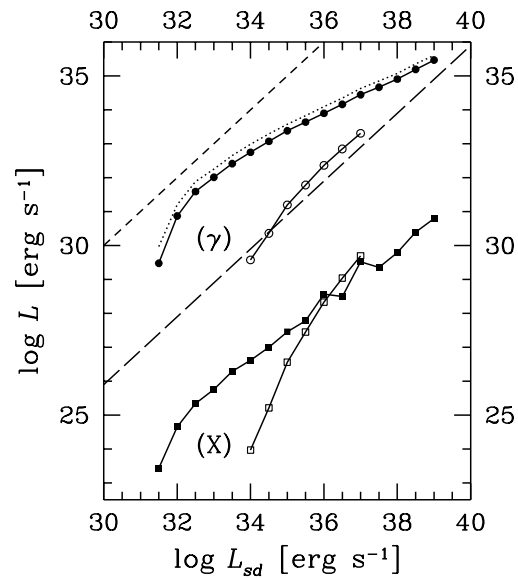}
\end{center}
\caption{
Evolution of high-energy, non-thermal luminosity across the spin-down
luminosity space. 
The long upper curve connecting filled dots is the track
of gamma-ray luminosity $L_\gamma(>100\keV)$, and the long lower curve connecting
filled squares is the track
of X-ray luminosity $L_X(0.1 - 10\keV)$,
both calculated according to eq.\ref{Lgx}, for a pulsar with $B_{\rm pc}=10^{12}\G$.
[The dotted line is the track of $L_{particles}$ (eq.\ref{Lp}), which in RD98 was identified
with $L_\gamma$.] \hfill\break 
The short upper curve with open circles is 
for gamma-ray luminosity $L_\gamma(>100\keV)$, and the short lower curve with
open squares is for X-ray luminosity $L_X(0.1 - 10\keV)$,
calculated according to eq.\ref{Lgx}, for $B_{\rm pc}=10^{9}\G$.
The short-dashed line marks $L = L_{\rm sd}$.
The long-dashed line, marking the empirical relation of Becker \& Tr{\"u}mper (1997)
rewritten for $\Omega_X = 1$~sr, has been added for reference.
The fluctuations of $L_X(0.1 - 10\keV)$ (for $10^{12}\G$ only) between $10^{36}\ergs$
and $10^{37}\ergs$ in \edot, are of numerical origin: in this range of \edot~
the low-energy tail of the synchrotron component crosses the curvature component
just around $10\keV$, and any fluctuations due to Monte Carlo treatment of 
cutting off synchrotron spectra influence the level of the total spectrum.
}
\label{fig:fig2}
\end{figure}

Extremely low values of relative energy output in X-ray photons and gamma-ray photons per
particle
imply equally low values of \Lx/\Lg~ (between $10^{-6}$ and $10^{-4}$).
Fig.2 presents \Lg~ and \Lx~ calculated according to eqs.\ref{Lgx}, \ref{yield}, and \ref{Lp}
as a function of \edot.
It shows unambiguously, that some pulsars from the list of X-ray and gamma-ray source
do not fit the picture: 
The Crab pulsar, for which the observed ratio of \Lx/\Lg~ is about 0.47, remains certainly a challenge.
Also among millisecond pulsars, there is a `Crab-like' example, which
does not fit the model. B1821-24 observed
with {\it ASCA} (Saito et al. 1997) reveals clear, double-peaked pulsations resembling
those of the Crab. Apart of the disagreement between the observed spectral slope ($\sim 1.9$)
and the predicted one ($\sim 1.5$), the predicted ratio of \Lx/\Lg~ 
$\sim 2 \times 10^{-4}$ (we took \Lx~ for 1sr) 
would make \Lg~ to exceed \edot. Or, alternatively, using the upper limit for 
the gamma-ray luminosity, $L_\gamma < 0.028 \, L_{\rm sd}$
(Nel et al. 1996), the ``observed" ratio \Lx/\Lg~ is $> 2.5 \times 10^{-3}$, in disagreement with
predictions.

However, among pulsars there are also strong sources of steady X-ray emission, with no trace of pulsations.
B1706-44, a strong gamma-ray pulsar detected with {\it EGRET}, with
$L_\gamma = 2.5 \times 10^{34}\ergs$ (Thompson et al. 1996)
shows no pulsations in X-rays. Its unpulsed, non-thermal X-ray emission has
the luminosity $L_X = 1.3\times 10^{33}\ergs$ (Becker et al. 1993).
Apparently, pulsed X-rays of B1706-44,
which we predict to reach the luminosity of
$\sim 3\times 10^{29}\ergs$, are dwarfed by nearby nebular steady emission.

\section{Conclusions}
Recent results from {\it ROSAT} and {\it ASCA}
suggest that the X-ray spectra of pulsars may be a superposition
of thermal 
and non-thermal components (see e.g. Zavlin \& Pavlov 1997 for discussion). 
The existence of these two components is in general expected in both
classes of rival models (polar-cap 
and outer-gap models; e.g. Arons 1981 and Cheng, Ho \& Ruderman 1986, respectively) 
and the thermal component results from 
polar cap heating by inflowing particles.  
However, clear empirical relations between thermal and non-thermal components 
are yet to be determined. The latest review 
on the status of pulsar' X-ray properties by Becker \& Tr{\"u}mper (1997) favoures pure non-thermal
spectral models (with four exceptions of additional thermal components due to
initial cooling of the star itself). 

Within the framework of polar-cap models with magnetospheric activity
induced by curvature radiation of beam particles we have calculated
broad-band photon spectra of non-thermal origin, for both classical
and millisecond pulsars. 
This non-thermal emission is a superposition of curvature and synchrotron radiation.
It is expected to be pulsed, like its high-energy
extension - the gamma-ray emission. 
Our objective was to estimate spectral properties and
the level of this emission in the energy domain of X-rays. 

Our calculations show, that for beam particles
with monoenergetic source function,
the properties of photon spectra in the combined energy range 
for {\it ROSAT} and {\it ASCA}
(0.1 - 10 keV) depend primarily on the magnetic field strength at the polar cap. Millisecond pulsars,
with dipolar magnetic 
fields of the order of $10^9\G$, have X-ray spectra dominated by synchrotron emission. On the other
hand,
pulsars with $B$ of the order of $10^{12}\G$ have X-ray spectra almost exclusively
due to low-energy tail of curvature emission.
However, the spectrum of
the curvature component breaks already at $150 \MeV$ regardless the pulsar parameters,
and neither its slope nor level in the X-ray domain is compatible with phase-averaged
spectra of pulsed X-ray emission inferred from observations.

Detailed properties of the broad-band spectrum, depend
upon the location(s) of cyclotron turnover energy 
$\epsilon_{\rm ct} = \hbar {e B \over m_{\rm e} c} /\sin \psi$ of the synchrotron emission. 
Unlike in outer-gap models, the available
range of pitch angles $\psi$ is rather narrow and confined to low values
($\sin \psi < 0.01/\sqrt P$). 
We find that for classical pulsars ($10^{12}\G$) the upper limit for turnover energy
$\epsilon_{\rm ct}$  occurs already at $\sim 1 \MeV$. Below this energy, the level
of synchrotron spectrum
gradually decreases, and around $\sim 10 \keV$ the 
curvature component takes over.
Unless the spectrum of primary beam particles is very steep 
($N_{\rm b} \propto \gamma_{\rm b}^{-4}$)
and extends down to Lorentz factor $\gamma_{\rm b} \approx 10^5$,  
the X-ray spectrum below $\sim 10 \keV$ is   
a power-law with photon index
$\alpha_{\rm ph} = 2/3$. Therefore, the model is not able
to explain the non-thermal X-ray spectra with $\alpha_{\rm ph} \simeq 1.5 \div 2$
inferred from 
observations 
(Becker \& Tr{\"u}mper 1997).\hfill\break 
The expected luminosity $L_{\rm X}$ of
pulsed component
is a negligible fraction of gamma-ray luminosity
$L_\gamma$,
and for $L_{\rm sd} \geq 10^{33}\ergs$
an approximate relation 
\begin{equation}
\log L_{\rm X} \approx -1.5 + 0.83\, \log L_{\rm sd}
\end{equation} 
follows from Fig.2.
The situation is qualitatively different for millisecond pulsars.
The X-ray spectra are dominated now by synchrotron components, with a photon index
$\alpha_{\rm ph} \simeq 3/2$, which extends
down to $\epsilon_{\rm ct} \sim 100 \eV$ before breaking sharply into
$\alpha_{\rm ph} \simeq -1$.
The expected $L_{\rm X}$ is still low with respect to
$L_\gamma$, though not as low as for classical pulsars.
For $L_{\rm sd} \geq 10^{34}\ergs$ we find  
\begin{equation}
\log L_{\rm X} \approx -40.3 + 1.90\, \log L_{\rm sd}.
\end{equation} 

We conclude that 
spectral properties and fluxes of pulsed non-thermal X-ray
emission of some objects, like the Crab or
the millisecond pulsar B1821-24, pose a real challenge to
the subclass of polar-cap models based on curvature and 
synchrotron radiation alone. 

\section*{ACKNOWLEDGEMENTS}
This work has been financed by the KBN grant 2P03D-00911.
BR acknowledges discussions with K.S. Cheng, J. Gil, and 
W. Klu\'zniak. 
We thank the anonymous referee for useful suggestions
which helped to clarify the paper.


\begin{thebibliography}{}

\bibitem[]{}Arons J., 1981, ApJ, 248, 1099
\bibitem[]{}Becker W., 1995, Ph.D. Thesis (LMU Munich) 
\bibitem[]{}Becker W., Brazier K.T.S., Tr{\"u}mper J., 1993, A\&A, 273, 421
\bibitem[]{}Becker W., Tr{\"u}mper J., 1997, A\&A, 326, 682 
\bibitem[]{}Cheng K.S., Gil J., Zhang L., 1998, ApJ, 493, L35
\bibitem[]{}Cheng K.S., Ho C., Ruderman M.A., 1986, ApJ, 300, 500
\bibitem[]{}Daugherty J.K., Harding A.K., 1982, ApJ, 252, 337 
\bibitem[]{}Daugherty J.K., Harding A.K., 1996, ApJ, 458, 278 
\bibitem[]{}Dermer C.D., Sturner S.J., 1994, ApJ, 420, L75 
\bibitem[]{}Harding A.K., Preece, R.T., 1987, ApJ, 319, 939
\bibitem[]{}Manning R.A., Willmore A.P., 1994, MNRAS, 266, 635  
\bibitem[]{}Nel H.I. et al., 1996, ApJ, 465, 898 
\bibitem[]{}O'Dell S.L., Sartori L., 1970, ApJ, 161, L63 
\bibitem[Rudak \& Dyks 1998]{RD98}Rudak B., Dyks J., 1998, MNRAS, 295, 337 [RD98]
\bibitem[]{}Saito Y., 1998 in Shibata S., Sato M., eds.
Proc. Neutron Stars and Pulsars. Universal Academy Press, Tokyo, in press
\bibitem[]{}Saito Y., Kawai N., Kamae T., Shibata S., Dotani T., Kulkarni S.R.,
1997, ApJ, 477, L37
\bibitem[]{}Thompson D.J. et al. 1996, ApJ, 465, 385
\bibitem[]{}Thompson D.J., Harding A.K.H., Hermsen W., Ulmer M.P.,
1997, in Dermer C.D., Strickman M.S., Kurfess, J.D., eds,
AIP Conf. Proc. 410, 4th Compton Symposium. AIP, New York, in press
\bibitem[]{}Wang F.Y.-H., Halpern J.P., 1997, ApJ, 482, L159
\bibitem[]{}Wang F.Y.-H., Ruderman M., Halpern J.P., Zhu T., 1997, preprint (astro-ph/9711283)
\bibitem[]{}Zhang L., Cheng K.S., 1997, ApJ, 487, 370
\bibitem[]{}Zhavlin V.E., Pavlov G.G., 1998, A\&A, 329, 583

\end{thebibliography}
\end{document}